\documentclass[12pt]{article}
 \usepackage[dvips]{graphicx}
 \usepackage[dvips]{graphics}
 \usepackage{amsmath}
 \usepackage{enumerate}
 \usepackage{psfrag}
 \newcommand{\newc}{\newcommand}
 \newc{\ra}{\rightarrow}
 \newc{\lra}{\leftrightarrow}
 \newc{\beq}{\begin{equation}}
 \newc{\eeq}{\end{equation}}

\begin{document}

 \begin{center}
 {\LARGE \bf
 Multi-toroidal configurations as equilibrium-flow eigenstates}

 \vspace{2mm}

 {\large G. Poulipoulis$^\dag$\footnote{me00584@cc.uoi.gr},
 G. N. Throumoulopoulos$^\dag$\footnote{gthroum@cc.uoi.gr}, H.
 Tasso$^\star$\footnote{het@ipp.mpg.de}}

 \vspace*{1mm}

 $^\dag${\it University of Ioannina, Association Euratom - Hellenic
 Republic,\\ \vspace{-1mm}
 Section of Theoretical Physics, GR 451 10 Ioannina, Greece}

 \vspace{1mm} \noindent $^\star${\it  Max-Planck-Institut f\"{u}r
 Plasmaphysik, Euratom Association,\\ \vspace{-1mm}
 D-85748 Garching, Germany }
 \end{center}

 \begin{center}
 {\bf Abstract}
 \end{center}
 Equilibrium eigenstates of an axisymmetric
 magnetically confined plasma with toroidal flow
 are investigated by means of
 exact solutions of the ideal magnetohydrodynamic equations.
 The study includes "compressible" flows with constant
 temperature, $T(\psi)$,  but varying density on magnetic surfaces and
 incompressible ones with constant density, $\varrho(\psi)$, but
 varying temperature thereon  (the function $\psi(R,z)$  is the poloidal magnetic flux-function with
 $(R,z,\phi)$ cylindrical coordinates). These variations are necessary for the existence of
 tokamak steady states with flow.
 The "compressible" and incompressible solutions are associated
 with the
 ansatz $\omega^2/T\propto M_0=\mbox{constant}$ and $d(\varrho\omega^2)/d\psi=A\psi$, respectively, where
 $\omega(\psi)$ is the
 rotation frequency. In both cases eigenfunctions of the form $\psi_{\ell n}={\cal Z}_\ell(z){\cal R}_n(R)$
 ($\ell, n = 1,2,\ldots$) describe configurations with $\ell \times n$ magnetic axes.
 Owing to the flow, the respective eigenvalues can be considered in two alternative ways:
 either as  flow eigenvalues, $M_0^\ell$ and $A^\ell$, which depend on a
 pressure  parameter, $P_0$, or as pressure eigenvalues, $P_0^\ell$,
 which  depend
 on the flow parameters $M_0$ or  $A$. In the context of the latter
 consideration when the flow parameters are varied  continuously there are transition
 points, $(M_0)_m$ and $A_m$ ($m=1,2,\ldots$),
 at which an additional magnetic axis is formed. This flow-caused change in magnetic topology
 is possible solely in the presence of toroidicity
 because in the limit of infinite aspect ratio the axial flow does
 not appear in the equilibrium equation. Also, the lower the aspect ratio
 the smaller $(M_0)_m$ and $A_m$.  In addition, the effects of the flow and
 the aspect ratio on the Shafranov shift are evaluated.
 \vspace{0.6cm}\\

 \noindent
 {\large \bf I. Introduction} \vspace{0.1cm} \\

 Over the last decades it has been established  experimentally and theoretically  that the flow affects
 the confinement properties of magnetically confined plasmas.
 In particular,  the flow  and especially
 the flow shear play a role in the formation of edge transport barriers
 (L-H transition) as well as of Internal Transport Barriers (ITBs),
 two enhanced confinement modes in tokamaks (e.g. see Ref. \cite{Te} and
 Refs. cited therein). Also, the majority of the  advanced tokamak scenarios include flow.
 The ITBs usually are associated with reversed magnetic shear profiles.
 In addition, a possible magnetic topology
 of static (no-flow) equilibria  with reversed current density in the
 core region  proposed in Ref. \cite{MaMe} consists of  multitoroidal configurations
 having non-nested magnetic surfaces.

 Magnetohydrodynamic (MHD) equilibrium equations for axisymmetric magnetically confined plasmas with (a)
 isothermal magnetic surfaces
 and toroidal flow and (b) incompressible flow of arbitrary direction
 have been obtained in Ref. \cite{MaPe} and \cite{TaTh98}, respectively.
 Although purely toroidal axisymmetric flows
 are inherent incompressible because of symmetry, the former equilibria can
 be regarded as "compressible" in the sense that, alike the latter ones,
 the density varies on magnetic surfaces. Respective exact solutions were
 constructed in Refs. \cite{MaPe,ClFa,ThPa} and \cite{TaTh98,SiTh,ThPo}
 and the impact of flow on certain equilibrium characteristics was examined therein.
 In particular, in. Ref. \cite{SiTh} we extended the well known
 Solov\'ev solution \cite{Sh,So} to unbounded incompressible plasmas and found
 that the flow and its shear can change the magnetic topology thus resulting
 in a variety of novel configurations of astrophysical and laboratory
 concern.

 The aim of the present study is to examine the possible impact of the flow on the magnetic topology
 for equilibria relevant to plasmas of fusion devices.
 The study includes equilibria with both "compressible" and incompressible toroidal
 flows. The main conclusion is that the flow in conjunction with  toroidicity can change the magnetic structure
 by the formation of addition magnetic axes. The role of the toroidicity is
 important, i.e. this formation is not possible in the limit of infinite aspect ratio.

 The outline of the report  is as follows. A  derivation of the "compressible" and incompressible
 equilibrium equations  in a unified manner is first reviewed and  respective exact
 solutions are presented in Sec. II. In Sec. III equilibrium eigenstates of
 a magnetically confined plasma surrounded by a  boundary of
 rectangular cross-section and arbitrary aspect ratio are constructed.
 On the basis of these eigenstates the impact of the  flow  on
 the magnetic topology is then studied in conjunction with the role of toroidicity.
 The effects of the flow and the aspect ratio on the Shafranov shift are
 examined in Sec. IV. Finally the conclusions are summarized in section V.\vspace{0.1cm}\\

 \noindent
 {\large \bf II. Equilibrium equations and solutions}\\

 The ideal axisymmetric MHD equilibrium equations for the cases of
 "compressible" and incompressible toroidal flows are reviewed in
 this Section. In particular, a unified derivation is given   without adopting from
 the beginning relevant energy equations or equations of states.
 They will be specified when necessary later. This rather detailed
 presentation aims at making the  discussion in the subsequent sections
 tangible, particularly as concerns the role of toroidicity.

 The starting equations
 written in standard notation and convenient units are the following:
 \begin{eqnarray}
 {\bf \nabla}\cdot (\varrho {\bf v})=0 \label{1} \\
 \varrho ({\bf v}\cdot {\bf
 \nabla}){\bf v}={\bf J}\times{\bf B}-{\bf \nabla}P
 \label{2} \\
 {\bf \nabla}\times{\bf E}=0 \label{3} \\
 {\bf \nabla}\times{\bf B}={\bf J} \label{4} \\
 {\bf \nabla}\cdot{\bf B}=0 \label{5} \\
 {\bf E}+{\bf v}\times{\bf B}=0 \label{6} \\
 \mbox{ An energy equation or equation of state } \label{7}
 \end{eqnarray}
 For axisymmetric magnetically confined plasmas
 with  toroidal flow the divergence-free magnetic field and
 mass flow can be written in terms of the scalar functions $\psi (R,z)$,
 $I(R,z)$ and $K(R,z)$ as
 \begin{eqnarray}
 {\bf B}=I{\bf \nabla}\phi +{\bf \nabla}\phi\times {\bf \nabla}\psi
 \label{8} \\
 \varrho {\bf v}=K{\bf \nabla}\phi. \label{9}
 \end{eqnarray}
 The toroidal current density is  then put, by Amp$\acute{e}$re's law,
 in the form
 \beq
 {\bf J}={\bf \Delta}^*\psi {\bf \nabla}\phi -{\bf \nabla}\phi\times
 {\bf \nabla}I
 \label{10}
 \eeq
 Here ($R, \phi, z$) are cylindrical coordinates
 with $z$ corresponding to the axis of symmetry, $\psi$ labels the
 magnetic surfaces and ${\bf \Delta}^*$ is the elliptic operator
 defined as $R^2{\bf \nabla}\cdot ({\bf \nabla}/R^2)$.

 By projecting the momentum equation  (\ref{2}) and Ohm's law
 (\ref{6}) along the toroidal direction, the magnetic field,  and
 perpendicular to the magnetic surfaces some integrals are identified as flux
 functions, i.e. functions
 constant on magnetic surfaces, and Eqs. (\ref{1}-{6}) are reduced to simpler
 ones.  In particular, the  $\nabla \phi$-component of (\ref{2})
  yields
 \beq
 {\bf \nabla}\phi\cdot ({\bf \nabla}\psi\times{\bf \nabla}I)=0
 \label{11}
 \eeq
 implying that $I=I(\psi )$. Also,  expressing the
 electric field in terms of the electrostatic potential, ${\bf E}=-{\bf \nabla} \Phi$,
  the component  of Ohm's law
 along $\bf B$ leads to
 \beq
 {\bf B}\cdot{\bf \nabla}\Phi =0.
 \label{12}
 \eeq
 Eq. (\ref{12}) implies that $\Phi=\Phi(\psi)$, viz.   $\bf E$ is perpendicular
 to  magnetic surfaces.
 One additional flux function is obtained by the projection of
 Ohm's law along $\nabla \psi$:
 \beq
 (\frac{d\Phi}{d\psi}-\frac{K}{\varrho R^2} )\cdot\left|{\bf
 \nabla}\psi\right|^2 =0
 \label{13}
 \eeq
 Therefore, the quantity
 \beq
 \frac{K}{\varrho R^2}\equiv\omega= \frac{d\Phi}{d\psi},
 \label{14}
 \eeq
 identified as the
 rotation frequency,
 is a flux function $\omega=\omega (\psi)$.

 With the aid of equations (\ref{11})-(\ref{14}) the components of
 equation (\ref{2}) along ${\bf B}$ and $\nabla \psi$
  respectively yield
 \begin{eqnarray}
 \bigg[\frac{{\bf \nabla}P}{\varrho}-{\bf
 \nabla}\Big(\frac{\omega^2R^2}{2}\Big)\bigg]\cdot{\bf B}=0
 \label{15} \\
 \big[{\bf \Delta}^*\psi+II'\big]|{\bf \nabla}\psi|^2+R^2\bigg[{\bf
 \nabla}P-\varrho\omega^2{\bf
 \nabla}\Big(\frac{R^2}{2}\Big)\bigg]\cdot{\bf \nabla}\psi=0,
 \label{16}
 \end{eqnarray}
 where the prime denotes differentiation with respect to $\psi$.

 In order to reduce Eqs. (\ref{15}) and (\ref{16}) further  an energy
 equation or an equation of state is necessary. Owing to the large heat
 conduction along $\bf B$,  isothermal magnetic surfaces, $T=T(\psi)$, is an
  appropriate equation of state for fusion plasmas. In
 this case employing the ideal gas law, $P=\lambda\varrho T$,
 integration of  (\ref{15}) yields
 \beq
 P=P_s(\psi)\exp{\Big(\frac{\omega^2R^2}{2\lambda T}\Big)}
 \label{19}
 \eeq
 where  $P_s(\psi)$ is the pressure
 in the absence of flow. In the presence of flow the pressure and
 therefore for $T=T(\psi)$ the density are   in general  not constant on magnetic
 surfaces, thus giving  rise to "compressibility".

 With the aid of (\ref{19}), Eq. (\ref{16}) leads to the final
 "compressible" equation
 \beq
 {\bf
 \Delta}^*\psi+II'+R^2\bigg[P_s'+P_s\frac{R^2}{2}\Big(\frac{\omega^2}{\lambda
 T}\Big)'\bigg]\exp{\Big(\frac{\omega^2R^2}{2\lambda T}\Big)}=0.
 \label{20}
 \eeq
 Eq. (\ref{20}) was originally obtained in Ref. \cite{MaPe}.

 An alternative equation of state is incompressibility:
 \beq
 {\bf \nabla}\cdot{\bf v}=0.
 \label{21}
 \eeq
 Consequently, (\ref{1}) implies that the density is a flux function,
 $\varrho=\varrho(\psi)$,  and integration of  (\ref{15}) yields
 \beq
 P=P_s(\psi)+\frac{R^2\varrho\omega^2}{2}.
 \label{22}
 \eeq
 Eq. (\ref{16}) then reduces to
 \beq
 {\bf \Delta}^*\psi+II'+R^2P_s'+\frac{R^4}{2}(\varrho\omega^2)'=0.
 \label{23}
 \eeq
 This is a particular form of the axisymmetric equilibrium equation
 for incompressible flow of arbitrary direction obtained in Ref.
 \cite{TaTh98}.

 Both equations (\ref{20}) and (\ref{23}) contain four
 flux-functions, three out of which, i.e. $P_s$, $I$ and $\omega$,
 are common. The fourth function is $T$ for the "compressible" equation
 and $\varrho$ for the incompressible one. For vanishing flow (\ref{20})
 and (\ref{23}) reduce to the Grad-Shafranov equation. The flow term in
 (\ref{20}) depends on $\omega$ and $\varrho$ through $\omega^2/\lambda T$
 and its $\psi$-derivative (shear) while solely the shear of the flow term
 $\varrho \omega^2$ appears in (\ref{23}).

 Linearized forms of Eqs. (\ref{20}) and (\ref{23}) in connection with  appropriate
 assignments of the free flux functions they contain can be solved
 analytically. In the present study we will employ exact solutions as follows.\\

 \noindent
 {\em ''Compressible'' flow}\\
 The ansatz used to linearize Eq. (\ref{20}) is \cite{ClFa}
 \cite{ThPa}
 \begin{eqnarray}
 I^2=I_0^2+I_1^2\psi^2 \nonumber \\
 P_s=2P_0\psi^2 \label{ans:1} \\
 \frac{\omega^2}{\lambda T}=\frac{\gamma M_0^2}{R_0^2}=\mbox{constant} \nonumber
 \end{eqnarray}
 Here,   $I_0/R$ is the toroidal vacuum field, the parameter $I_1$
 describes the magnetic properties of the plasma;  $P_0$, $\gamma$,  and $M_0$
 are a pressure parameter, the ratio of  specific heats, and the Mach number with
 respect to the  sound speed at a reference point ($z=0, R=R_0$) with $R_0$ to be specified later.
 Note that the toroidal current density profile  can vanish on the plasma
 boundary via (\ref{8}).

 Eq. (\ref{20}) then
 has a separable solution, ${\cal R}(R){\cal Z}(z)$, when the
 constant of separation is equal to $I_1$. For configurations symmetric
 with respect to mid-plane $z=0$ this  solution is written in the form
 \beq
 \psi (x,y)=C_1\bigg[J_0\Big(\frac{2\, \tau{\sqrt{e^
 {\frac{\gamma\,{M_0}^2\,x^2}{2}}}}}{\gamma\, {M_0}^2}\Big)+C_2Y_0\Big(\frac{2\, \tau{\sqrt{e^
 {\frac{\gamma\,{M_0}^2\,x^2}{2}}}}}{\gamma\,
 {M_0}^2}\Big)\bigg]\cos{(I_1 y)},
 \label{24}
 \eeq
 where
 $x=R/R_0$ and $y=z/R_0$; $J_0$ and $Y_0$ are
 zeroth-order Bessel functions of first- and second-kind,
 respectively; and
 $\tau^2 \equiv 4P_0R_0^4$.\\

 \noindent
 {\em Incompressible flow}\\
 In this case the ansatz employed to linearize (\ref{23}) is
 \cite{ThPo}
 \begin{eqnarray}
 I^2=I_0^2+I_1^2\psi^2 \nonumber \\
 P_s=2P_0\psi^2 \label{ans:2} \\
 \Big[\frac{K^2}{\varrho R^4}\Big]'=2A\psi \nonumber
 \end{eqnarray}
 A remarkable difference of this choice compared to (\ref{ans:1}) is that the flow term  has
 non-zero shear [$(\varrho\omega^2)^\prime \neq 0$] and consequently $A$ can be positive or negative.
 Also, note that, unlike $M_0$ in (\ref{ans:1}), $A$ is dimensional.

 A separable solution is now expressed in terms of  the first- and
 second-kind Airy functions, $A_i$ and $B_i$, as \cite{ThPo}
 \begin{eqnarray}
 \psi (x,y)=C_1\Bigg[
 Ai\bigg(\Big(\frac{AR_0}{4}\Big)^{-2/3}\Big(\frac{AR_0^6}{4}x^2-P_1R_0^4\Big)\bigg) \nonumber \\
 +C_2 Bi\bigg(\Big(\frac{AR_0}{4}\Big)^{-2/3}\Big(\frac{AR_0^6}{4}x^2-P_1R_0^4\Big)\bigg)\Bigg]\cos{(I_1
 y)}.
 \label{25}
 \end{eqnarray}

 \noindent
 {\large \bf III. Multitoroidal eigenstates associated with the flow}\\

\begin{figure}[!h]
 \begin{center}
 \includegraphics[scale=0.8]{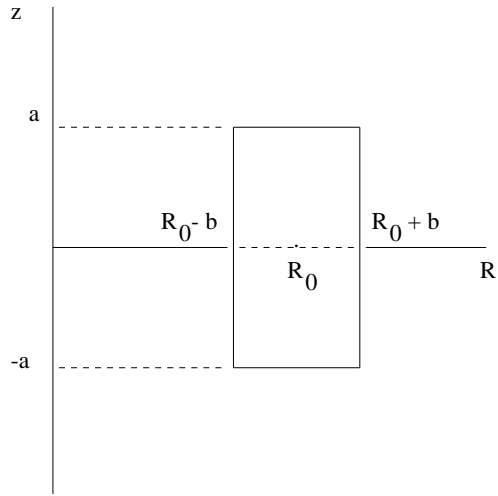}
 \caption{The  cross section of the plasma boundary. The aspect ratio $\alpha$ is defined as
 $R_0/b$.}
 \label{fig:1}
 \end{center}
 \end{figure}

 In connection with solutions (\ref{24}) and (\ref{25}) we are interested in the steady
 states of a tokamak the plasma of which is bounded by a
 conducting wall of rectangular cross-section, as shown in Fig.
 \ref{fig:1}.
 In addition, we assume that the plasma boundary coincides
 with the outermost magnetic surface; accordingly, the function $\psi$ should satisfy
 the following boundary conditions
 \beq
 \psi(y_\pm) = 0
 \label{25a}
 \eeq
 and
 \beq
 \psi(x_\pm) = 0,
 \label{25b}
 \eeq
 where $y_\pm=\pm \mbox{a}/R_0$ and $x_\pm=1\pm b/R_0$. The equilibrium becomes then a boundary-value problem.
 Eigenstates can be determined by imposing conditions (\ref{25a})
 and (\ref{25b}) directly to  solutions  (\ref{24}) and (\ref{25}).
 Specifically, (\ref{25a}) applied to  the $z$-dependent part of the solutions
 yields the eigenvalues
 \beq
 I_1^\ell = \frac{1}{\mbox{a}}\left(\ell \pi -\frac{\pi}{2}\right),\ \
 \ell=1,2,\ldots
 \label{25c}
 \eeq
 for the quantity $I_1$ which is related to the poloidal current function
 $I(\psi)$.
 The respective eigenfunctions are associated with
 configurations possessing $\ell$ magnetic axes parallel to the axis of symmetry.
 Condition (\ref{25b}) is pertinent to the $R$-dependent part of
 the solution. Owing to the flow  this part
 contains the parameter $M_0$ in the compressible case and $A$ in the
 incompressible one in addition to the  pressure parameter $P_0$.
 (To make  further discussion easier we introduce the  symbol $F$
 representing either $M_0$ or $A$; this is particularly
 convenient to formulate results which are independent of
 "compressibility".) Thus,  condition $\ref{25b})$ can determine
 eigenvalues  depending on the parameter
 $P_0$ which remains free, $F^n(P_0)$ ($n=1,2,3,\ldots$), or vice versa, pressure eigenvalues $P_0^n(F)$
 with $F$ being free.
 This parametric dependence makes the spectrum of the eigenvalues
 broader in comparison with the static one.
 The eigenvalues depend also on the geometrical quantities $R_0$ and
 $b$ but not on $\mbox{a}$
 (see Fig. \ref{fig:1}); thus, the results to follow are independent of elongation
 $\mbox{a}/b$. The other parameters $C_1$ and $C_2$
 contained in (\ref{23}) and (\ref{24}) are adapted to normalize $\psi$
 with respect to the magnetic axis and to satisfy the boundary
 condition (\ref{25b}) respectively.
 Also, in the rest of the report dimensionless values of $A$ will be given
 normalized  with respect to 1 Kgr/(m$^ 7$ T$^2$ s$^2$).
 The eigenvalues $F^n(P_0)$ and $P_0^n(F)$ can be calculated  numerically.
 For  a fixed value of $P_0$,  $F^n$
 satisfy for all $n$ the inequality $F^{n+1} > F^n$. A similar relation is satisfied by
 $P_0^n$ for a given value of $F$. The respective eigenfunctions are connected to
 configurations having $n$ magnetic axes perpendicular to the axis of symmetry.
 Therefore, the total equilibrium eigenfunctions $\psi_{\ell n}={\cal Z}_\ell(z){\cal R}_n(R)$
 describe multitoroidal configurations  having $\ell\times n$
 magnetic axes. For example, a  static doublet configuration,
 $\psi_{12}$, possessing two magnetic axes parallel to $z$-axis was studied in Ref.
 \cite{Ma}.

\begin{figure}[!h]
 \begin{center}
 \psfrag{R}{x}
 \psfrag{z}{y}
 \includegraphics[scale=0.8, trim=200 100 200 100 clip]{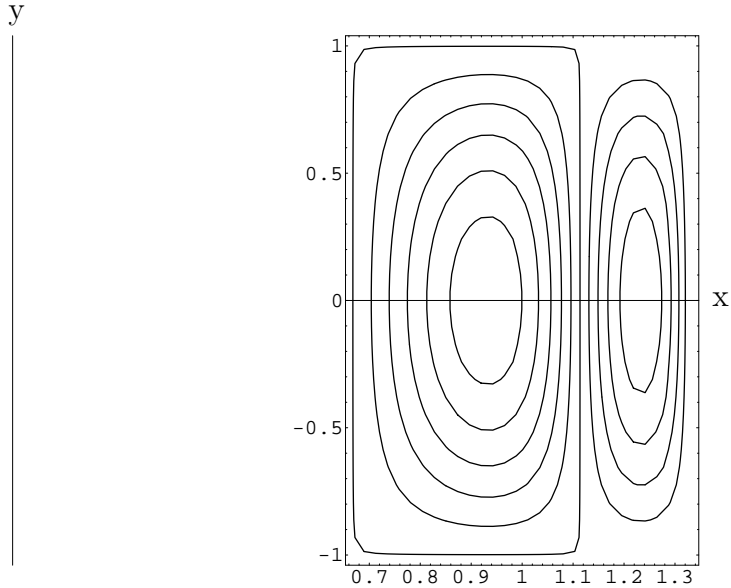}
 \caption{The double-magnetic-axis toroidal "compressible" configuration
 in connection with the eigenfunction  $\psi_{12}$
 with eigenvalue $M_0^2=1.692$ for aspect ratio $\alpha=3$.}
 \label{fig:2}
 \end{center}
 \end{figure}

\begin{figure}[!h]
 \begin{center}
 \psfrag{R}{x}
 \psfrag{z}{y}
 \psfrag{a}{(a)}
 \psfrag{b}{(b)}
 \psfrag{c}{(c)}
 \psfrag{d}{(d)}
 \psfrag{e}{(e)}
 \psfrag{f}{(f)}
 \includegraphics[scale=1]{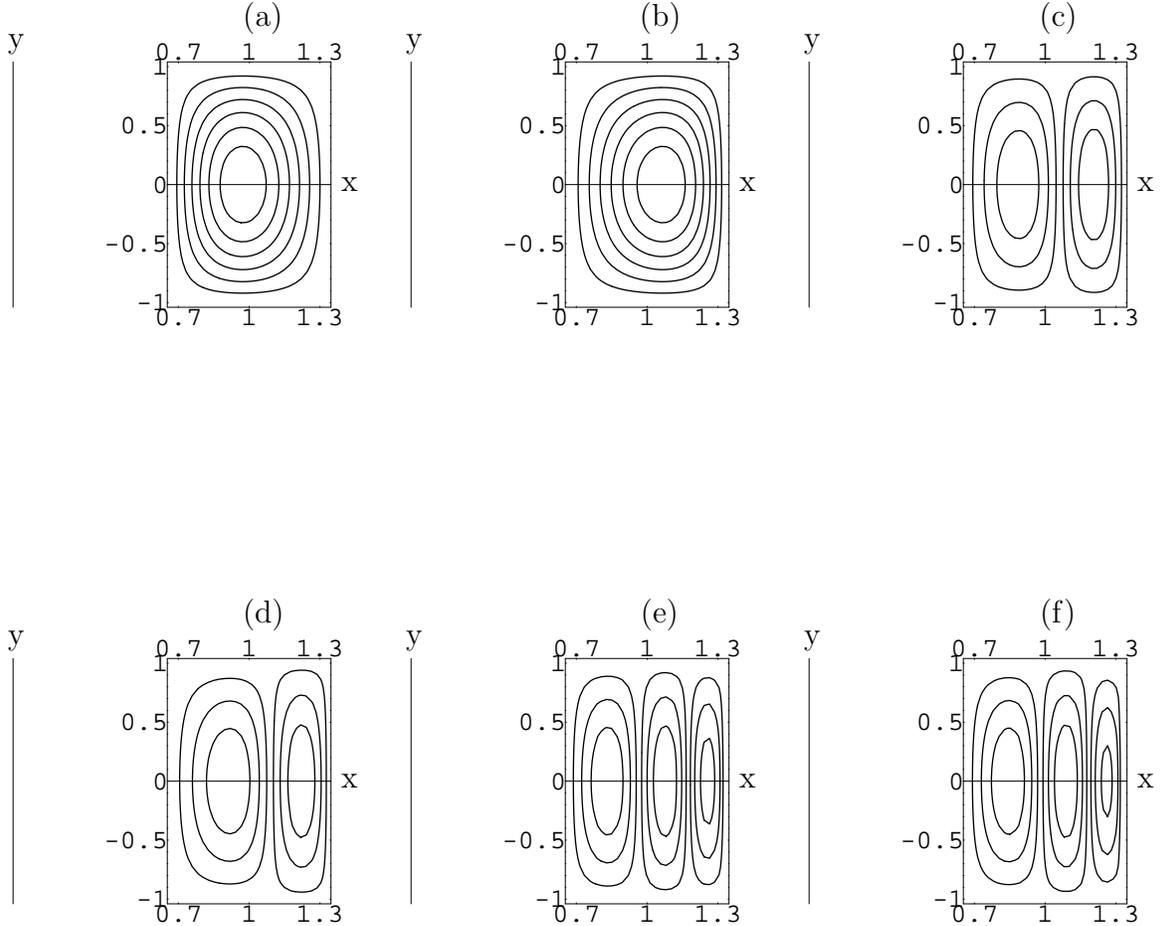}
 \caption{A sequence of graphs showing the quasi-static "evolution" of the
 configuration as the incompressible flow parameter $A$ is decreased for aspect ratio
 $\alpha=3$. The values of $A$  for the individual plots are
 (a) $A=0.09$, (b) $A=-0.01$, (c) $A=-0.02$, (d) $A=-0.1$, (e)
 $A=-0.122$, (f) $A=-0.2$.}
 \label{fig:3}
 \end{center}
 \end{figure}

 Henceforth, for the sake of simplicity we will restrict the study to eigenfunctions
 $\psi_{1n}$ describing multitoroidal configurations with $n$ magnetic axes along the mid-plane
 $z=0$. As an example the $\psi_{12}$-configuration  for
 "compressible" flow is shown in (Fig. \ref{fig:2}).
 For vanishing flow, tokamak multitoroidal configurations of this kind
 were investigated in Ref. \cite{MaMe} in connection with hole current
 density profiles which can reverse in the core region; current reversal is possible
 because the configurations have non-nested  magnetic surfaces.
 It is noted that for static equilibria  only pressure eigenvalues
 are possible. In the presence of flow we examined eigenstates with
 the lowest of the pressure eigenvalues $P_0^n(F)$ by varying continuously the
 flow parameter $F$ starting from a value close to the first-order static one, $P_0^1(F\approx 0)$.
 It should be noted here that for incompressible flow, $A=0$ does
 not necessarily imply static equilibrium because of the non-zero shear [see Eq. (\ref{ans:2})].
 It turns out that there are transition points $F_m \ (m=1,2,\ldots )$ at which the
 configuration changes topology by the formation of an additional magnetic axis
 (The subscript $m$ here indicating  a transition point must not
 be mixed up with the superscript $n$ indicating the order of an eigenvalue).
 For incompressible flow this is shown in Fig. \ref{fig:3}.
 Specifically,  the singly toroidal configuration  \ref{fig:3}(a)
 has eigenfunction $\psi_{11}$ with eigenvalue $P_0^1(A=0.09)$. By varying the flow parameter the pressure
 (eigenvalue) decreases and the configuration is shifted outwards
 and is compressed in the outer region  while the pressure becomes
 lower (\ref{fig:3}(b)). Then, as the flow parameter reaches the
 first transition point $A_1=-0.01$ a second magnetic axis is
 formed in the outer region close to the boundary causing the first one
 to be shifted inwards (\ref{fig:3}(c)). This transition point is associated with pressure
 eigenvalue $P_0^2(A=A_1)$. By a further increase of the flow (lower negative values of $A$) the
 outer magnetic island becomes bigger and the whole configuration is
 shifted outwards up to  the next transition point $A_2=-0.1$  associated with
 pressure eigenvalue $P_0^3(A=A_2)$ (Fig. \ref{fig:3}(d))
 at which as before  a third magnetic axis is formed in the
 outer region; thus, the configuration consists of three magnetic islands
 (Fig. \ref{fig:3}(e)) and is shifted outwards as
 the flow increases even more  (Fig. \ref{fig:3}(f)). This procedure is carried on
 until the formation of $n$ magnetic axes and it is also possible for "compressible" flow.
 It can  approximately be regarded as a quasi-static "evolution" of the plasma due to the  flow variation
 through flow-depended pressure equilibrium eigenstates.  Therefore,
 it is not possible for static equilibria. An animation showing
 this "evolution" is available in Ref \cite{mo}. Alternatively, varying continuously the pressure parameter
 $P_0$ one can find pressure transition points, $(P_0)_m$,
 associated with flow eigenvalues.

 For aspect ratio $\alpha=3$ and
 compressible flow the first transition velocity  from a singlet to a doublet configuration
 is of the order of $10^5$ m/s. Velocities of this order have been
 measured in tokamaks and therefore the change in  magnetic
 topology by the flow can be experimentally examined.

 It is emphasized that the change of the magnetic topology due to the
 flow is not possible in the limit of infinite aspect ratio  because
 in this limit the equilibrium equations do not contain the axial
 velocity regardless of "compressibility". Indeed, for a cylindrical
 plasma of arbitrary cross-section the  equations respective to (\ref{15})
 and (\ref{16}) read
 \begin{eqnarray}
 {\bf B}\cdot{\bf \nabla}P=0
 \label{17} \\
 \nabla^2\psi+\Big(P+\frac{B_z^2}{2}\Big)'=0.
 \label{18}
 \end{eqnarray}
 Eqs. (\ref{17}) and (\ref{18}) follow respectively from (16) and (17) of Ref \cite{ThTa97} for vanishing
 poloidal velocity ($F^\prime=0$).
 Also, note that the pressure becomes a flux-function.

 In addition, toroidicity affects qualitatively the eigenvalues and
 the transition points. We examined this impact by varying the aspect
 ratio $\alpha$ and found the following results:
 \begin{enumerate}
 \item
 The eigenvalues $P_0^n$ and $F^n$ ("compressible and
 incompressible)
 become lower as the aspect ratio decreases. For example, for
 $\alpha=3$ and $\alpha=2$ one finds $M_0^2=2.8$ and $M_0^2=2.3$,
 respectively.
 \item
 The smaller $\alpha$ the lower the transition
 Mach numbers $(M_0)_m$  and $A_m$ (for any $m$). For example,
 for $\alpha=3$ and $\alpha=2$ the respective
 first-transition Mach numbers  are $(M_0)_1=1.692$ and $(M_0)_1=1.338$ and the first-transition
 incompressible points are $A_1=-0.083$ and $A_1=-0.448$.
 \end{enumerate}
 The "compressible"
 transition  velocities are found to be in general supersonic. Experimental supersonic toroidal
 velocities have been reported in Refs. \cite{Bay} and \cite{Gua}. Owing
 to the above mentioned dependence of the transition points on $\alpha$, however,
 it is possible  to have subsonic transitions for
 appropriate low values of $\alpha$.
 Thus, in this case the transitions may be realized easier
 in spherical tokamaks; the minimum subsonic value of the first-transition point $(M_0)_1$  is 0.62
 and corresponds to a compact toroid ($\alpha=1$). \\

 \noindent
 {\large \bf IV. Shafranov shift}
  \vspace{0.1cm} \\

 We  evaluated the impact of the flow and the aspect
 ratio on the Shafranov shift of  eigenstates with a single magnetic axis.
 (It is reminded that the Shafranov shift is defined as the
 displacement of the magnetic axis with respect to the geometrical
 center of the configuration, $\Delta x\equiv x_{m.a.}-1$.)
 The results are summarized as follows.

 \begin{enumerate}
 \item As $M_0$ increases or $A$ decreases the Shafranov shift increases.
 This increase for "compressible and incompressible equilibria can be seen in
 Tables \ref{tab:1} and \ref{tab:2}, respectively. It is interesting to note that for large
 positive values of $A$ the Shafranov shift can
 become negative. As an example for $\alpha=3$ and $A=0.09$ the shift is $-0.0274$.
 This means that the magnetic surfaces in this case are shifted
 inward. An inward shift of magnetic surfaces
 associated with poloidal flow in quasi-isodymanic equilibria
 was reported in Ref. \cite{HuGu}. Also, suppression of the
 Shafranov shift by a properly shaped toroidal rotation profile was
 found in Ref. \cite{IlPo}.

 \begin{table}[!h]
 \begin{center}
 \begin{tabular}{|c|c|}
 \hline
 $M$  & Shafranov Shift \\
 \hline
 0.1 &  0.054 \\
 \hline
 0.4 & 0.058 \\
 \hline
 0.6 & 0.063 \\
 \hline
 \end{tabular}
 \caption{ The Shafranov shift, $\Delta x\equiv x_{m.a.}-1$,  for various values of the
 Mach number ($M_0$) and  aspect ratio $\alpha=3$.}
 \label{tab:1}
 \end{center}
 \end{table}

 \begin{table}[!h]
 \begin{center}
 \begin{tabular}{|c|c|}
 \hline
 A & Shafranov Shift \\
 \hline
 0.010 & 0.045 \\
 \hline
 0.006 &  0.049 \\
 \hline
 -0.001 & 0.055 \\
 \hline
 \end{tabular}
 \caption{The Shafranov shift, $\Delta x\equiv x_{m.a.}-1$, for various values of the
 incompressible-flow parameter $A$ and aspect ratio $\alpha=3$.}
 \label{tab:2}
 \end{center}
 \end{table}

 \item The lower the aspect ratio the larger the Shafranov shift.
 This is shown in tables \ref{tab:3} and \ref{tab:4} for two different values of
 the pressure parameter $P_0$.

  \begin{table}[!h]
 \begin{center}
 \begin{tabular}{|c|c|}
 \hline
 Aspect ratio & Shafranov shift \\
 \hline
 3 & 0.092 \\
 \hline
 2 & 0.150 \\
 \hline
 1.5 & 0.209 \\
 \hline
 \end{tabular}
 \caption{The Shafarnov shift, $\Delta x\equiv x_{m.a.}-1$, for $P_0=12$ kPa and various values of the aspect
 ratio for the "compressible" case.}
 \label{tab:3}
 \end{center}
 \end{table}

 \begin{table}[!h]
 \begin{center}
 \begin{tabular}{|c|c|}
 \hline
 Aspect ratio &  Shafranov shift \\
 \hline
 3 & 0.053 \\
 \hline
 2 & 0.140 \\
 \hline
 1 & 0.500 \\
 \hline
 \end{tabular}
 \caption{The Shafarnov shift, $\Delta x\equiv x_{m.a.}-1$, for  $P_0=110$ kPa and various values of the aspect
 ratio for the incompressible case.}
 \label{tab:4}
 \end{center}
 \end{table}

 \end{enumerate}

 \newpage
 \noindent
 {\large   \bf V. Summary and Conclusions}\\

 Equilibrium eigenstates of a  magnetically confined plasma with toroidal flow
 surrounded by a boundary of rectangular
 cross-section have been investigated within the framework of ideal MHD theory.
 "Compressible" flows associated with  uniform temperature but varying density on magnetic
 surfaces and incompressible ones with uniform density but
 varying temperature thereon have been examined on the basis of respective reduced equations
 [(\ref{20}) and (\ref{23})] and exact solutions [(\ref{24}) and
(\ref{25})]. The flow effect on the magnetic topology of the
eigenstates has been examined by means of the
 parameters $M_0$ and $A$ associated with the quantities
 $\omega^2/T$ in the "compressible" and $\varrho \omega^2$ in
 incompressible case, respectively. The exact "compressible" solutions
 considered are shearless $[(\omega^2/( T))^\prime=0]$
 while the incompressible ones have non-zero shear [$(\varrho \omega^2)^\prime\neq 0$].

 Owing to the flow one can consider either pressure eigenvalues,
 $(P_{0})^n\ n=1,2,\ldots$ with the flow parameter $F$ being free
 ($F$ represents either $M_0$ or $A$) or alternatively flow eigenvalues
 $F^n$ with free $P_0$. For fixed $F$ in the former case and fixed $P_0$
 in the latter one, the eigenvalues satisfy  the relations $P_0^{n+1}>P_0^n$
 and $F^{n+1}>F^n$, respectively. The respective eigenfunctions for the
 poloidal magnetic flux-function $\psi$ can describe multitoroidal configurations
 with $n$ magnetic axes located on the mid-plane $z=0$. When $M_0$ is increased
 or $A$ is decreased continuously there are transition points $F_m$
 ($m=1,2,\ldots$) at which an additional magnetic axis appears. Alternatively,
 by varying continuously the pressure parameter $P_0$ there are transition
 points $(P_0)_m$  associated with flow eigenvalues at which an additional
 magnetic axis is formed. This change in magnetic topology, possible only
 in the presence of flow, is crucially related to toroidicity because in
 the limit of infinite aspect ratio the equilibrium equations are flow
 independent. The above mentioned "flow-triggered" transitions can be
 approximately viewed as a quasi-static "evolution" of the plasma by
 continuous flow variation through pressure eigenstates or alternatively
 by continuous pressure variation through flow eigenstates. The transition
 points have the following dependence on the aspect ratio $\alpha$: the
 lower $\alpha$  the smaller $(M_0)_m$ in the "compressible" case and $A_m$
 in the incompressible one.

 Also, we have examined the impact of the flow and the aspect
 ratio $\alpha$ (as a measure of the  toroidicity) on the Shafranov
 shift. The results show that the shift (a) increases  as $M_0$ take
 larger values and $A$ smaller ones and (b) increases as $\alpha$ takes
 lower values. Furthermore, for large positive values of $A$ the shift
 can become negative.

 \end{document}